\documentclass[aps,prd,showkeys,showpacs,twocolumn,superscriptaddress]{revtex4}

\usepackage{amsfonts}
\usepackage{amsmath}
\usepackage{latexsym}
\usepackage{amssymb}
\usepackage[mathcal]{euscript}
\usepackage{graphicx}

\newcommand{\ul}{\underline}

\newcommand{\trace}{\textrm{Tr}}

\newcommand{\ov}{\overrightarrow}
\newtheorem{Lem}{Lemma}
\newtheorem{Prop}{Proposition}

\begin{document}

\title{\begin{center}
A Note on the correspondence between \\
Qubit Quantum Operations and Special Relativity
\end{center}
}

\author{Pablo Arrighi}
\email{pja35@cam.ac.uk} \affiliation{Computer Laboratory,
University of Cambridge, 15 JJ Thomson Avenue, Cambridge CB3 0FD,
U.K. }
\author{Christophe Patricot \setcounter{footnote}{5}}
\email{cep29@cam.ac.uk} \affiliation{ DAMTP,
University of Cambridge, Centre for Mathematical Sciences,\\
Wilberforce Road, Cambridge CB3 0WA, U.K.}



\begin{abstract}
We exploit a well-known isomorphism between complex hermitian
$2\times 2$ matrices and $\mathbb{R}^4$, which yields a convenient
real vector representation of qubit states. Because these do not
need to be normalized we find that they map onto a Minkowskian
future cone in $\mathbb{E}^{1,3}$, whose vertical cross-sections are
nothing but Bloch spheres. Pure states are represented by
light-like vectors, unitary operations correspond to special
orthogonal transforms about the axis of the cone, positive
operations correspond to pure Lorentz boosts. We formalize the
equivalence between the generalized measurement formalism on qubit
states and the Lorentz transformations of special relativity, or more precisely elements of the
restricted Lorentz group together with future-directed null boosts. The note ends
with a discussion of the equivalence and some of its
possible consequences.
\end{abstract}

\maketitle

\section{Preliminaries and Geometrical Setting}
This paper may be viewed as a complement to conal representations of
quantum states \cite{us}. This section reproduces some of the
material in a concise manner, in an attempt to make the
presentation self-contained.

The state of a two dimensional quantum system (a qubit) is an
element of $\textrm{Herm}_2^+(\mathbb{C})$ the set of $2\times 2$
positive complex matrices \cite{Nielsen}. Traditionally one tends
to consider normalized states only, i.e. unit trace
$\textrm{Herm}_2^+(\mathbb{C})$ matrices (density matrices). Yet
relaxing this condition has a clear physical meaning and we will
often do so in this note. The most general evolution a qubit state
may undergo is a generalized measurement (the only extra feature
Kraus operators allow is the possibility to ignore one's knowledge
of some measurement outcomes). These are described by a finite set
$\{M_m\}$ of $2\times 2$ complex matrices satisfying $\sum_{m}
M_m^{\dagger}M_m=\mathbb{I}$. If we let $E_m=M_m^{\dagger}M_m$ we
have that $\sum_{m}E_m=\mathbb{I}$,
$E_m\in\textrm{Herm}^+_2(\mathbb{C})$ and $M_m=U_m\sqrt{E_m}$
using the polar decomposition. Applied upon a density matrix
$\rho$, the generalized measurement $\{M_m\}$ yields outcome $m$
with probability $p(m)=\trace(E_m\rho)$, in which case the
post-measurement state is given by $\rho'_m=(1/ \trace(E_m
\rho))(M_m \rho M_m^{\dagger})$. We shall call $\rho_m= M_m \rho
M_m^{\dagger}\in \textrm{Herm}^+_2(\mathbb{C})$ the
\emph{unrescaled} post-measurement state. Note that the
generalized measurement formalism can be viewed as arising when
the system is first coupled to an ancilla (through a unitary
operation), which then gets measured projectively and discarded.
This paper takes the more axiomatic view on generalized quantum
measurements.

Let $\{\sigma_{\mu}\}_{\mu = 0...3}$ designate the set of the Pauli
matrices $\mathbb{I},\mathbf{X},\mathbf{Y}$ and $\mathbf{Z}$.
These form a Hilbert-Schmidt orthogonal basis of $2 \times 2$
hermitian matrices, that is $\forall \;\mu,\nu \quad
\trace(\sigma_{\mu}\sigma_{\nu})= 2\delta_{\mu\nu}$ with $\delta$
the Kronecker delta. Thus any matrix $A\in
\textrm{Herm}_2(\mathbb{C})$ decomposes on this basis as
\begin{align*}
A&=(1/2)\big(\trace(A)\mathbb{I}+\trace(A\sigma_i)\sigma_i\big)=(1/2)\trace(A\sigma_{\mu})\sigma_{\mu}
\end{align*}
Notice that throughout this article latin indices run from $1$ to
$3$, greek indices from $0$ to $3$, and repeated indices are
summed unless specified. Letting
$\ul{A}_{\mu}=\trace(A\sigma_{\mu})$, we shall call $\ul A$ the
vector $(\ul{A}_{\mu})\in\mathbb{R}^{4}$ while
$\overrightarrow{A}=(\ul A_i)$ will designate the restricted
vector in $\mathbb{R}^{3}$. Note that the coordinate map
\begin{align*}
\phi: \textrm{Herm}_2(\mathbb{C})&\to \mathbb{R}^{4} \\
        A & \mapsto \ul A
\end{align*}
is an isometric isomorphism, in the sense that
\begin{align}
\label{isometry} \trace(AB)&=\frac{1}{2}\ul{A}.\ul{B}\equiv
\frac{1}{2}\ul{A}_{\mu}\ul{B}_{\mu}
\end{align}

\begin{Lem}
The cone of positive hermitian matrices
$\textrm{Herm}_2^+(\mathbb{C})$ is isomorphic to the following
cone of revolution in $\mathbb{R}^4$:
\begin{align*}
\Gamma&=\{(\lambda_\mu) \in \mathbb{R}^4 \,/\, \lambda_0^2
-\sum_{i=1}^3 \lambda_i^2 \geq 0, \lambda_0 \geq 0\}
\end{align*}
Generalized pure states lie on the boundary of $\Gamma$.
\end{Lem}
\textbf{Proof}: Let $A\in \textrm{Herm}_2(\mathbb{C})$. Its eigenvalues  are given by
$\lambda_\pm=\frac{1}{2}(\ul A_0 \pm \sqrt{\ul A_i \ul A_i})$.
$A$ is positive if and only if $\lambda_+ \lambda_- \geq 0$
and $\lambda_+ + \lambda_- \geq 0$. This is equivalent to:
\begin{align*}
\eta_{\mu\nu}\ul A_{\mu} \ul A_{\nu} \geq 0 \quad \textrm{and} \quad \ul A_0
\geq 0
\end{align*}
with $\eta_{\mu \nu}=\textrm{Diag}(1,-1,-1,-1)$.  Moreover $A$ is
generalized pure when one of its eigenvalues is zero, which is
equivalent to $\eta_{\mu\nu}\ul A_{\mu} \ul A_{\nu} = 0$.{$\quad\Box$}\\
Thus the generalized (not necessarily normalized) density matrices
of a qubit cover the whole Minkowskian future-light-cone in
$\mathbb{E}^{1,3}$. Taking a vertical cross-section of the cone
is equivalent to fixing the trace $\ul A_0$ of the density matrix,
which might be thought of physically as the \emph{overall
probability of occurrence} for the state. By doing so we are left
with only the spin degrees of freedom along $\mathbf{X}$,
$\mathbf{Y}$ and $\mathbf{Z}$, and therefore each vertical
cross-section is a Bloch sphere with radius $\ul A_{0}$.


Where the use of Clifford algebras is encountered such a
representation is not totally uncommon. We think, for instance, of
the community of geometric algebras \cite{Havel}. Furthermore
$\phi^{-1}$ is precisely the isomorphism used to define Dirac
spinors \cite{pages jaunes} in Quantum Field Theories. For
$n$-dimensional extensions of the representation we refer the
reader to \cite{us} and \cite{Zanardi} .

We now consider the map $\psi$ from $2\times 2$ complex matrices to
endomorphisms of $\mathbb{R}^4$ given by:
\begin{align*}
\psi: A\mapsto\phi\circ Ad_A \circ \phi^{-1}
\end{align*}
i.e. $\psi(A)$ is the $4 \times 4$ real matrix taking a vector
$\ul{\rho}$ into $\ul{A \rho A^{\dagger}}$. Notice that
$\psi(AB)=\psi(A)\psi(B)$. Amongst the standard results
\cite{Nielsen} we have that $\psi(U)$, with $U$ unitary, is a
special orthogonal transform about the axis of revolution of the
cone $\Gamma$. Indeed without loss of generality one can assume
$\textrm{det}(U)=1$, and so the special unitary matrix can be written as:
\begin{align}
\label{rotation}
U=\cos(\frac{\theta}{2})\mathbb{I}-i\sin(\frac{\theta}{2})(\ov{n_k}
\sigma_k)&=e^{-i\frac{\theta}{2}\;\ov{n_k} \sigma_k}\\
\textrm{and has image:}\quad \psi(U)&=\left(\begin{array}{ccc} 1&&0\\
0&&R_{\theta}(\ov{n})\end{array}\right)\nonumber
\end{align}
Here $R_{\theta}(\ov{n})$ denotes the real rotation by an angle
$\theta$ around the normalized axis $\ov{n}$ (to happen in the Bloch sphere).
Alternatively one may use the expression $\psi(U)_{\mu
\nu}=(1/2)\trace(U\sigma_{\nu}U^\dagger\sigma_{\mu})$. The next
formulae are not well-known.

\begin{Lem}
Let $\sqrt{E_m}$ be a matrix in $\textrm{Herm}_2^+(\mathbb{C})$,
with $\ul{\sqrt{E_m}}=[\alpha\;\beta\;\gamma\;\delta]$, and $E_m$
its square, with $\ul{E_m}=[a\;x\;y\;z]$. Then
\begin{align}
\psi(\sqrt{E_m})&=\frac{1}{4}\left(\begin{array}{cccc}
-X\!+\!2\alpha^2     &2\alpha\beta &2\alpha\gamma &2\alpha\delta\\
2\alpha\beta     &X\!+\!2\beta^2   &2\beta\gamma  &2\beta\delta\\
2\alpha\gamma    &2\beta\gamma &X\!+\!2\gamma^2   &2\gamma\delta\\
2\alpha\delta    &2\beta\delta &2\gamma\delta &X\!+\!2\delta^2
\end{array}\right)\nonumber\\
\label{psi em}&=\frac{1}{4}\left(\begin{array}{cccc}
2a               &2x                  &2y                  &2z\\
2x               &X\!+\!\frac{4x^2}{2a+X} &\frac{4xy}{2a+X}    &\frac{4xz}{2a+X}\\
2y               &\frac{4xy}{2a+X}    &X\!+\!\frac{4y^2}{2a+X} &\frac{4yz}{2a+X}\\
2z               &\frac{4xz}{2a+X}    &\frac{4yz}{2a+X}    &X\!+\!\frac{4z^2}{2a+X}\\
\end{array}\right)\\
\textrm{with}\quad
X&=\alpha^2-\beta^2-\gamma^2-\delta^2=2\sqrt{a^2-x^2-y^2-z^2}.\nonumber
\end{align}
\end{Lem}
\textbf{Proof:} $\psi(\sqrt{E_m})$ can be computed in terms of
$\ul{\sqrt{E_m}}$ using the following simple formula:
\begin{align*}
\psi(\sqrt{E_m})_{\mu
\mu'}&=(1/2)\ul{\sqrt{E_m}}_{\nu}\ul{\sqrt{E_m}}_{\nu'}\trace(\sigma_{\nu}\sigma_{\mu'}\sigma_{\nu'}\sigma_{\mu})
\end{align*}
This method requires lengthy calculations, subtler approaches are
discussed in \cite{us}. Now let
$\ul{\iota}=[1\;0\;0\;0]=(1/2)\phi(\mathbb{I})$ and observe that
\begin{align}
\psi(\sqrt{E_m})\ul{\iota}&\equiv\phi\circ Ad_{\sqrt{E_m}}\circ 
\phi^{-1}\ul{\iota}\nonumber\\
\label{to square} &=(1/2)\phi(\sqrt{E_m}\mathbb{I}\sqrt{E_m})\equiv
(1/2)\ul{E_m}
\end{align}
In other words, $(1/2)\ul{E_m}$ has as components  the first column of
$\psi(\sqrt{E_m})$. Thus we can now proceed to the substitutions which yield the
second form of $\psi(\sqrt{E_m})$. Finally the $X$ relation stems
from:
\begin{align}
\label{X relation}\eta_{\mu
\nu}\ul{\sqrt{E_m}}_{\mu}\ul{\sqrt{E_m}}_{\nu}&=4\,\textrm{det}(\sqrt{E_m})\\
&=4\,\sqrt{\textrm{det}(E_m)}=2\sqrt{\eta_{\mu
\nu}\ul{E_m}_{\mu}\ul{E_m}_{\nu}}\quad \Box \nonumber
\end{align}

\section{Quantum operations as Lorentz transforms and vice-versa}
\label{formal correspondence}
We begin by showing that elements of a generalized measurement act
on a qubit either as rescaled restricted Lorentz transformations
or as rescaled future-directed null boosts. Then we show that the
reverse is also true. Remember that a Lorentz transform $L\equiv
L^{\mu}_{\nu}$ is called restricted if it is proper
($\textrm{det}\,L=1$) and orthochronous ($L^0_0>0$). We will show
that such an $L$ decomposes uniquely into the product of a proper spatial
rotation and a pure (timelike future-directed velocity) boost. We like
to think of null velocity boosts as limiting cases of restricted boosts, or effectively as elements of the topological boundary of
the restricted Lorentz group, but they need to be \emph{rescaled} to yield a finite
linear transform. We shall call these (rescaled) future-directed null 
boosts. They are singular transforms. It turns out the rescaling introduced defines a
natural unifying way of thinking about Lorentz transforms and null
boosts.

If $\ul{E_m}=[a, \;x, \; y,\; z ]$  corresponds to one particular
measurement element $E_m=M_m^{\dagger}M_m$, we shall call
$\ul{V_m}$ the vector of coordinates 
$(\ul{V_m}_{\mu})=(\frac{1}{2} \eta_{\mu \nu}\ul{E_m}_{\nu})$, i.e. $\ul{V_m}=[a/2,\; -x/2,\; -y/2,\; -z/2]$.
Then $\ul{v_m}=2\ul{V_m}/a$ is the corresponding normalized vector 
and $\ov{v_m}=[-x/a ,\; -y/a, \; -z/a]$ can be thought of as a
three vector velocity, whose norm is defined as usual: $v_m=(\ov{v_m}.\ov{v_m})^{1/2}$.

\begin{Prop}
\label{measurement as lorentz}
Let $\{M_m\}=\{U_m\sqrt{E_m}\}$ be a generalized measurement on a
qubit, with $U_m$ unitary and $\sqrt{E_m}$ positive. Then for all $m$
such that $E_m$ is not projective, we have:
\begin{align}
\label{lorentz up to scale}
\psi(M_m)&=\sqrt{\eta_{\mu \nu}\ul{V_m}_{\mu}\ul{V_m}_{\nu}}R_mL(\ul{v_m})
\end{align}
where $R_m=\psi(U_m)$ is a proper rotation about the axis of the
cone and $L(\ul{v_m})$ is a pure restricted Lorentz boost of normalized velocity
$\ul{v_m}$. Thus $\psi(M_m)$ is a restricted Lorentz transform up
to a (strictly positive) scalar. Similarly, if $E_m$ is
projective,
$\psi(M_m)=(a/2)R_mL(\ul{v_m})$, where $L(\ul{v_m})$ is a
rescaled pure future-directed null boost of null velocity
$\ul{v_m}$. 
\end{Prop}
\textbf{Proof:} First recall that
$\psi(M_m)=\psi(U_m)\psi(\sqrt{E_m})$, and by (\ref{rotation}),
$\psi(U_m)$ is a special orthogonal transformation about the axis
of the cone, so a restricted Lorentz transform. Suppose $\ul{E_m}$
(hence $\ul{v_m}$) 
timelike future-directed. Letting $\gamma\equiv2a/X$ in (\ref{psi
em}) and using the definition of $\ov{v_m}$, we get:
\begin{equation}
\label{pure boost}
\frac{4}{X}\psi(\sqrt{E_m})=\left(\begin{array}{ccc}
\gamma & & -\gamma \ov{v_m}^T \\
-\gamma \ov{v_m} & & \mathbb{I}
+\frac{\gamma^2}{1+\gamma}\ov{v_m}\ov{v_m}^T \end{array}\right)\equiv
L(\ul{v_m})
\end{equation}
As $\gamma=1/\sqrt{1-v_m^2}$, $L(\ul{v_m})$ is precisely a pure Lorentz boost of velocity
$\ul{v_m}$ (see \cite{pages jaunes} for example). Since $\ul{v_m}$ is timelike future-directed, $\psi(M_m)$ is a restricted
Lorentz transform up to the factor $X/4=(1/2)\sqrt{\eta_{\mu
\nu}\ul{E_m}_{\mu}\ul{E_m}_{\nu}}=\sqrt{\eta_{\mu
\nu}\ul{V_m}_{\mu}\ul{V_m}_{\nu}}$.\\
Now, when $\ul{E_m}$ is null ($E_m$ projective) this factor
vanishes and $\gamma$ becomes infinite. Nevertheless one can write
(\ref{psi em})
for $X=0$ as
\begin{equation}
\label{null boost}
\frac{2}{a}\psi(\sqrt{E_m})=\left(\begin{array}{ccc}
1 & & -\ov{v_m}^T \\
-\ov{v_m} & &  \ov{v_m}\ov{v_m}^T \end{array} \right)
\end{equation}
We can see that this is in fact a pure null boost \emph{rescaled} by a
factor $\gamma^{-1}$. Indeed, when $v_m\to1$ the right-hand-side of (\ref{pure boost}) becomes
\begin{align*}
L^{null}(\ul{v_m})\sim \gamma\left(\begin{array}{ccc}
1 & & -\ov{v_m}^T \\
-\ov{v_m} & &  \ov{v_m}\ov{v_m}^T \end{array} \right)
\end{align*}
and since $\frac{a}{2}=\gamma\sqrt{\eta_{\mu \nu}\ul{V_m}_{\mu}\ul{V_m}_{\nu}}$, 
we precisely get
\begin{displaymath}
\psi(\sqrt{E_m})\sim\sqrt{\eta_{\mu
\nu}\ul{V_m}_{\mu}\ul{V_m}_{\nu}}L^{null}(\ul{v_m})
\end{displaymath}
Here the Minkowski product vanishes and the \emph{unrescaled} pure null velocity boost
is infinite. Nevertheless rescaling $L^{null}(\ul{v_m})$ by the factor
$\gamma^{-1}$ yields the right-hand-side of (\ref{null boost}); thus 
$\psi(\sqrt{E_m})$ indeed corresponds to a  rescaled pure null boost, which of
course is not an element of the Lorentz group. $\quad$ $\Box$ 

As we said previously the natural rescaling by the Minkowski product precisely
corresponds to an appropriate rescaling of generalized Lorentz transforms bringing null
boosts to finite linear maps.  Formally the essence of this Proposition can be
thought of as a consequence of the Alexandrov-Zeeman theorems relating
the causality group (Lorentz group and dilatations) to the Minkowskian
causal structure, though this approach would not cover null
velocity boosts. Note that the rescaled pure null velocity boosts (right-hand-side of
(\ref{null boost})) are in fact proportional to projections on the null four vectors
$\ul{E_m}$. \\ 
Maybe the reader wonders here why the Lorentz pure boosts corresponding
to positive measurement elements $E_m$ are parametrized by $\ul{v_m}$
and not $\ul{E_m}$. However since $E_m$ is an operator acting on states
and not a state,
$\ul{E_m}$ is better thought of as a \emph{co-vector}, or element of
the \emph{dual} space,  in the same way as momenta are dual to
positions in usual Special Relativity. The (contravariant) vector
corresponding to $\ul{E_m}$ is precisely $2\ul{V_m}$, thus in the space of
states, and not operators, $E_m$ is represented by
$2\ul{V_m}$. The factor of two  was introduced merely for convenience. \\

The following relations suggest the Minkowski product of the state
vector of a qubit is an important quantum information theoretical quantity:
\begin{Prop}
\label{info prop}
Let $\{M_m\}$ be a generalized measurement, $\ul{\rho}$ a state vector
and $\psi(M_m)\ul{\rho}\equiv\ul{\rho_m}$ the unrescaled
post-measurement state vector if outcome
$m$ occurs. We have:
\begin{align}
\label{information}
\eta_{\mu \nu}\ul{\rho_m}_{\mu}\ul{\rho_m}_{\nu}&=\eta_{\mu
\nu}\ul{V_m}_{\mu}\ul{V_m}_{\nu}\eta_{\mu'
\nu'}\ul{\rho}_{\mu'}\ul{\rho}_{\nu'}\\
\label{invariant probability}
\ul{\rho_m}_0&=\eta_{\mu\nu}\ul{V_m}_{\mu}\ul{\rho}_{\nu} \\
\label{trace}
\eta_{\mu \nu}\ul{\rho}_{\mu}\ul{\rho}_{\nu}&=
2([\textrm{\emph{Tr}}(\rho)]^2-\textrm{\emph{Tr}}(\rho^2))
\end{align}  
\end{Prop}
\textbf{Proof:} We make use of the previous proposition. Equation
(\ref{lorentz up to scale}) implies
\begin{displaymath}
\eta_{\mu \nu}\ul{\rho_m}_{\mu}\ul{\rho_m}_{\nu}=\eta_{\mu
\nu}\ul{V_m}_{\mu}\ul{V_m}_{\nu}\eta_{\mu'
\nu'}(R_m L(\ul{v_m})\ul{\rho})_{\mu'}(R_m L(\ul{v_m})\ul{\rho})_{\nu'}
\end{displaymath}
and (\ref{information}) follows since $R_m L(\ul{v_m})$ is a Lorentz
transform. This relation remains true of course when $\ul{V_m}$ is
light-like ($E_m$ projective), since so is $\ul{\rho_m}$. (Purity relations
\cite{us}). \\
For the second equation note that $\ul{\rho_m}_0=\trace(E_m \rho)
=(1/2)\ul{E_m}.\ul{\rho}$, where the isometry (\ref{isometry}) was
applied. 
Introducing the definition of $\ul{V_m}$ in this last equation yields the
required result. \\
Equation (\ref{trace}) can be shown explicitly using the components of
$\rho$ and $\rho^2$,  but it seems more
interesting to use our isomorphism $\phi:\rho\to\ul{\rho}$. Consider the linear map on
$\mathbb{E}^{1,3}$, $\Lambda:(\ul\rho_{\mu})\to (\eta_{\nu
\mu}\ul\rho_{\nu})$ (musical isomorphism). Then
$\widetilde{\Lambda}:\rho \to \phi^{-1}\circ \Lambda \circ \phi
(\rho)$ is a linear map on $\textrm{Herm}_2(\mathbb{C})$. One finds
easily $\widetilde{\Lambda}(\rho)=(\trace\rho)\mathbb{I}-\rho$. Using
the fact that $\phi$ is an isometry (\ref{isometry}), we get
\begin{align*}
\eta_{\mu \nu}\ul{\rho}_{\mu}\ul{\rho}_{\nu}&\equiv
(\Lambda\ul\rho).\ul\rho=2\trace(\widetilde{\Lambda}(\rho)\rho) \\
&= 2([\trace\rho]^2-\trace(\rho^2)) \quad \quad \Box
\end{align*}
It seems interesting that this quantity, invariant under Lorentz
transforms on the state vector $\ul\rho$, in fact measures the
mixedness of qubit states: recall that a density matrix $\rho$ is
pure if and only if $\trace(\rho^2)=(\trace\rho)^2$. Not only is
purity preserved under a formal Lorentz boost, so is this notion of mixedness. Moreover this
quantity maps according to the simple relation (\ref{information})
under a generalized measurement. Note that since $\eta_{\mu
\nu}\ul{V_m}_{\mu}\ul{V_m}_{\nu}\leq 1$, the mixedness always
decreases given a measurement outcome. But (\ref{information}) and
(\ref{invariant probability}) suggest
much more: the mixedness of
post-measurement states and their probabilities are 
invariant if both the initial vector $\ul{\rho}$ and the measurement
vectors $\ul{V_m}$ 
are Lorentz transformed. However, the set of transformed measurement
vectors does not sum to the identity, and it is unclear how to interpret
it as a quantum measurement. In
section \ref{part three} we will discuss the way a boosted observer
perceives measurement probabilities, but without using the approach
equation (\ref{invariant probability}) might suggest. We now show that any Lorentz
transformation  can be thought of as an element of a generalized
measurement up to scale.

\begin{Prop}
\label{lorentz as measurement} Let L a restricted Lorentz
transform or a rescaled future-directed null boost of
$\mathbb{E}^{1,3}$. $L$ decomposes as $L=RL(\ul{v})$ where $R$ is
a proper Lorentz rotation and $L(\ul{v})$ a pure velocity boost,
rescaled when $\ul{v}$ is null. Then there
exits a particular element of a measurement scheme $\{M_m\}$,
$M_1$ say, such that for any qubit $\rho$,
\begin{equation}
L\ul\rho \propto{\psi(M_1)\ul\rho}
\end{equation}
Thus the effect of a Lorentz boost on a qubit can essentially be viewed as
applying a particular measurement element whose outcome occurs.
More precisely there exits a family of
such possible measurement elements $M(\lambda)=U\sqrt{E(\lambda)}$
defined by $U=U(R)$ as in (\ref{rotation}) and $\sqrt{E(\lambda)}$
satisfying the following:\\
If $L=RL(\ul{v})$ is a restricted Lorentz transform:
\begin{align*}
\ul{\sqrt{E(\lambda)}}&=(1+\sqrt{1-v^2})^{-1/2}[\lambda(1+\sqrt{1-v^2}),
-\lambda \ov{v}] \nonumber \\
&\textrm{with}\quad 0<\lambda \leq \sqrt{\frac{2}{1+v}} \nonumber
\end{align*}
while if $L=RL(\ul{v})$ is a rescaled future-directed null boost:
\begin{align*}
\ul{\sqrt{E(\lambda)}}&= [\lambda, \;-\lambda\ov{v}] \quad
\textrm{with} \quad 0<\lambda \leq1. \nonumber
\end{align*}

\end{Prop}
\textbf{Proof:} For completeness we first show the decomposition
of restricted Lorentz transforms $L$ into $L=RL(\ul{v})$ as above.
This relies on the well-known spinor representation of the
restricted Lorentz group, or the two-to-one group homomorphism
between unimodular $2\times 2$ complex matrices and restricted
Lorentz transforms (see \cite{pages jaunes} for example):
\begin{align*}
\psi: SL(2,\mathbb{C}) &\to SO(1,3)^+ \\
          A &\mapsto       \psi(A)\equiv \phi\circ Ad_A \circ \phi^{-1}
\end{align*}
Indeed as $Ad_A$ preserves the determinant and  $\phi$ is such
that for all $\rho \in \textrm{Herm}_2(\mathbb{C})$,
$\textrm{det}\,\rho=(1/4)\eta_{\mu
\nu}\ul\rho_{\mu}\ul\rho_{\nu}$, $\psi(A)$ preserves the Minkowski
product. The fact that $\psi(A)\in SO(1,3)^+$ and that $\psi$ is
two-to-one and onto can be checked explicitly. Let $L$ any
restricted Lorentz transform. There exits a unique $A\in
SL(2,\mathbb{C})$ such that $\psi(\pm A)=L$. Polar decompose $A$
into $A=U|A|$ with $U$ unitary and $|A|$ positive. ($U$ is in fact
special unitary and $|A|$ positive definite since
$\textrm{det}\,A=1$, and by unicity of the polar decomposition for
$A$ non-singular, $-A=(-U)|A|$). Applying Proposition
\ref{measurement as lorentz} to $|A|$ with
$\textrm{det}\;|A|^2=1$, $\psi(|A|)$ is a \emph{pure} restricted
Lorentz boost, thus $L=\psi(U)\psi(|A|)$ provides a decomposition.
Since
$\psi(U)=\psi(-U)$, this decomposition is unique. \\
Thus given $L=RL(\ul{v})$, with
$R$ a proper rotation and $L(\ul{v})$ a pure boost of future-directed
timelike velocity
$\ul{v}=[1,\;\ov{v}]$, we use Proposition \ref{measurement as lorentz}
to find $M=U\sqrt{E}$ such $\psi(M)\propto L$. $U=U(R)$ is given by
(\ref{rotation}) and we choose $\ul{E}=[1,\;-\ov{v}]$.\\
We then have to find $\lambda>0$ such that $\lambda M$ can be part
of a measurement scheme. This is equivalent to
$\lambda^2M^{\dagger}M$ positive (satisfied) and
$\mathbb{I}-\lambda^2M^{\dagger}M$ positive too. ($\lambda M$ and
$-\lambda M$ are equivalent in terms of measurement elements). With $\lambda M=U\sqrt{E(\lambda)}$, we
have 
\begin{displaymath}
\ul{E(\lambda)}=[\lambda^2, \; -\lambda^2\ov{v}],
\end{displaymath}
from which we
find $\ul{\sqrt{E(\lambda)}}$ using (\ref{to square}):
\begin{align*}
\ul{\sqrt{E(\lambda)}}&=(1+\sqrt{1-v^2})^{-1/2}[\lambda(1+\sqrt{1-v^2}),\;
-\lambda \ov{v}].
\end{align*}
Then requiring $\mathbb{I}-E(\lambda)$ positive is equivalent to
($\lambda >0$):
\begin{displaymath}
\lambda \leq \sqrt{\frac{2}{1+v}}
\end{displaymath}
Applying Proposition \ref{measurement as lorentz} we get:
\begin{displaymath}
\psi(M(\lambda))=\frac{\lambda^2}{2}\sqrt{1-v^2}RL(\ul{v})
\end{displaymath}
Thus for such $\lambda$ the measurement elements
$M(\lambda)=U\sqrt{E(\lambda)}$ are all possible measurements whose occurrence is
equivalent up to a factor  to the restricted Lorentz boost $L=RL(\ul{v})$.

Now let $L$ a rescaled future-directed null boost. As we have shown, any restricted Lorentz transform can
be decomposed into a product of a proper rotation and a boost of
time-like future-directed velocity. Future-directed null boosts are
just limits of these, and thus the rescaled null boosts $L$ may be assumed to be the
product of a rotation $R$ and a rescaled null pure boost $L(\ul{v})$ of type (\ref{null
boost}). The rotation can be dealt with as in the previous
case. Defining  $\ul{E}=[1 ,\; -\ov{v}]$ null future-directed,
we have $L(\ul{v})\propto \psi(\phi^{-1}(\ul{\sqrt{E}}))$. Then again we consider $\ul{E(\lambda)}=\lambda^2\ul{E}$
$(\lambda>0)$ such that
$\mathbb{I}-E(\lambda)$ is positive. This is equivalent to
$0<\lambda\leq1$, and using (\ref{to square}) we have
\begin{displaymath}
\ul{\sqrt{E(\lambda)}}= [\lambda, \;-\lambda\ov{v}]
\end{displaymath}
which gives  $\psi(\sqrt{M(\lambda)})=(\lambda^2/2)RL(\ul{v})$. $\quad \Box$ \\
Note that the scaling factor is always
less than 1, indeed less than $\sqrt{(1-v)/(1+v)}$ in the restricted case, and
$1/2$ in the null case. 

Overall we have shown that elements of generalized measurements on a qubit
are \emph{equivalent} to rescaled restricted or null Lorentz
transforms. Projective measurement elements are future-directed null boosts,
while mixed ones correspond to restricted Lorentz boosts. One can of
course think of these linear transforms as elements or limits of elements of the causality
group of $\mathbb{E}^{1,3}$.

\section{Discussion}
\label{part three} The following is a somewhat original discussion
of Propositions \ref{measurement as lorentz} to \ref{lorentz as
measurement}. Our formalism and its consequences suggest that qubit
states may be viewed as spatio-temporal objects, or indeed as
four-vectors of a Minkowski space-time. This differs only slightly
from the notion of spin as a spatial polarization direction, and thus
may apply to 2 dimensional quantum systems whose degrees of freedom
can be thought of as spacelike. We shall adopt this point of view from now, i.e consider
naively qubits as four-vectors, and analyse the physical
implications. \\
Let us begin by merely rephrasing the content of the
correspondence that was established in section \ref{formal
correspondence}. Suppose Alice proceeds to a generalized
measurement $\{M_m\}=\{U_m\sqrt{E_m}\}$, $\sum_m M_m^{\dag}
M_m =\mathbb{I}$ on a qubit density matrix $\rho$ ($\rho$ is unit
trace). With probability $p(m)=\trace{(E_m \rho)}$ this will yield
her a (non-normalized) post-measurement state $\rho_m=M_m\rho
M_m^{\dag}$. This rather common situation turns out to be
equivalent, according to Proposition \ref{measurement as lorentz},
to the following less usual scenario:

\emph{Scenario 1:} Suppose Alice is standing at the origin of an inertial frame of
Minkowski spacetime, contemplating the four-vector $\ul \rho$. Say
she gives herself a set of rotations $\{R_m\}$ and four-vectors
$\{\ul{V_m}\}$ such that $\sum_m \ul{V_m} =[1\;0 \; 0\; 0]$.
Now, with probability $p(m)=\eta_{\mu\nu}\ul{V_m}_{\mu}\ul{\rho}_{\nu}$ she
chooses to Lorentz boost herself up to velocity vector
$\ul{v_m}=\ul{V_m}/\ul{V_m}_0$, to rotate the
resulting space-frame by $R_m$ and to rescale her coordinates by a
factor of $\sqrt{\eta_{\mu\nu} \ul{V_m}_{\mu} \ul{V_m}_{\nu}}$ (we
are assuming $E_m$ is not projective). She then looks
back upon her object of contemplation and sees $\ul{\rho_m}$, the
unrescaled post-measurement state.
The case with $E_m$ projective is the limit of the previous one
when the boost vector $\ul{v_m}$ becomes null, and the rescaling
yields finiteness of the corresponding linear transform. \\
Therefore a quantum measurement can be thought of, up to scale, as the
observer taking a Lorentz boost relative to his or her qubit. Notice that
applying a second quantum measurement $\{N_n\}$ similarly
corresponds to the observer taking a second (successive) Lorentz
transformation at random amongst $\{L_n\}$, say. Thus qubit
quantum mechanics can easily be axiomatized within the mathematics
of special relativity, and pure measurement elements go
hand-in-hand with future-directed null boosts.

Difficulties are
prompt to arise when one seeks to equate a measurement interaction, in which the qubit is physically acted upon,  with a (somewhat
passive) coordinate transformation in Minkowski space-time: indeed the
latter is purely kinematical, thus reversible, whereas
the former usually implies a collapse of the state. In the
following scenario we dissociate one from the other. In other
words we consider special relativity and qubit quantum theory in
their most usual fashion, save for the fact that we continue to
interpret the spin a a four-vector.

\emph{Scenario 2:} Suppose Alice at the origin of an inertial frame of
Minkowski space, together with a qubit density matrix $\rho$ (unit
trace) which we think of as a (normalized) space-time
vector $\ul{\rho}$. If we consider the point of view of Bob as he
passes by in an inertial frame, this suggests that Bob sees a
boosted version of $\rho$, i.e. a state $\Lambda\ul{\rho}$. This
seemingly innocuous point raises an important issue however:
$\Lambda$ is not restricted to Bloch sphere rotations, and thus
may indeed not correspond to a unitary operation. To understand
its effect upon $\rho$ we must refer to Proposition \ref{lorentz
as measurement}: $\Lambda$ acts, up to a  factor, as a measurement
element $M_1$ whose outcome always happens, even though
$\trace(M_1\rho M_1^{\dagger})\neq 1$. Thus $\{M_1\}$ can be
thought of as a non trace-preserving quantum operation ($M_1
M_1^{\dagger}\neq \mathbb{I}$) which systematically occurs. We
shall let $\ul{\rho^{Bob}}\equiv\Lambda\ul{\rho} \propto{\ul{
M_1\rho M_1^{\dagger}}}$ and proceed to reassure the reader: such
a phenomenon would not violate the principle of relativity. Bob
does not \emph{make happen} a non trace-preserving quantum
operation on the qubit. The laws of quantum mechanics remain
exactly the same in every inertial frame: only the \emph{change of
observers}, or more precisely the way a boosted observer perceives
a non-boosted state, is a non-orthodox quantum operation. If Bob
were then to decelerate down to the speed of Alice, his
mathematical description of the qubit would return to be $\rho$
again. \\
Now suppose Alice measures $\rho$ under a generalized measurement
$\{N_n\}$. The probability associated with the transition from
$\rho$ to $\rho_n$ is given by
$p(n)\equiv\trace(N_n^{\dagger}N_n\rho)/\trace(\rho)=\ul{\rho_n}_0$,
as usual when $\ul{\rho}$ is normalized. As Bob passes, he sees
the initial state $\ul{\rho^{Bob}}=\Lambda\ul\rho$, and the
post-measurement states $\ul{\rho^{Bob}_n}=\Lambda\ul{\rho_n}$.
Remember that the probability associated to a state is simply
given by the first component of its vector representation.
Assuming $\Lambda$ is a pure boost of non-null normalized velocity
$\ul{v(\Lambda)}$, we get:
\begin{displaymath}
\label{proba boost}
p^{Bob}(n)\equiv \frac{\trace(\rho_n^{Bob})}{\trace(\rho^{Bob})}=\frac{\ul{\rho_n^{Bob}}_0}{\ul{\rho^{Bob}}_0}=
\frac{p(n)-\ov{v(\Lambda)}.\ov{\rho_n}}{1-\ov{v(\Lambda)}.\ov{\rho}}
\geq 0
\end{displaymath}
In other words the probabilities associated with the transitions from
$\ul{\rho}$ to $\ul{\rho_n}$, in the same way as lengths of objects, are not invariant under a change of
observer.
Thus if one believes probabilities are absolute quantities
independent of notions of space and time, one must abandon trying to interpret
the qubit as a 4-vector. \\
Otherwise, the
notion of probability as a physical quantity needs to be redefined
($\sum_n p(n)$ is not conserved, as the probability of a state
transforms just like the time-component of a four vector). The
idea is disturbing, and certainly worth comparing with the
contraction of any spatial object (a ruler, say) under a Lorentz
boost. As he passes by Bob will see Alice's $20cm$ ruler shrunk
down to $15cm$. But what we now have is that if Alice's quantum
ruler has half a chance of being $22cm$ long, and another half
chance of measuring $18cm$, it may well turn out that Bob instead
perceives a quantum ruler of length $17cm$ with probability a
third, and $14cm$ two third of
the times.

Allowing the Lorentz boosts $\Lambda$ to act on $\ul{\rho}$ as on
space-time vectors thus seems a
radical departure from Quantum Field Theories in Minkowski space, where the approach
is to seek \emph{unitary} representations of the Poincar\'e 
 group, i.e. the full Lorentz group together with translations. However,
Poincar\'e invariance (see \cite{Wald} for example) does not require any
given state of a theory to transform unitarily under
a change of \emph{observer}: for any two inertial observers Alice and Bob, it
requires the existence, given any state of the theory possibly
measured by Alice in her frame, of
another state of the theory measured by Bob in his frame, such that
the statistics of their measurement outcomes  on their respective states be the same.   
In this sense, the action of a particular Poincar\'e transform  on a
state in Quantum Field Theory corresponds
to a change of \emph{inertial frame}: it maps a given solution for
an inertial family of observers to
another \emph{equivalent} solution for another family of observers, hence it   
simply cannot change the measurement statistics. Our second scenario
does not involve a change of inertial frame, but just a change of
observer. It is true that nonetheless, Alice's non-boosted qubit
viewed by a boosted observer Bob, though not necessarily unitarily equivalent to the same
non-boosted state viewed by Alice, should be an
admissible state of the theory which could be measured by Bob to yield
measurement statistics with the usual properties. We are not in this case, since in
scenario 2, Bob is not performing a quantum operation on Alice's qubit.   
Note also that in the formalism developed above, pure states, whether
viewed in their inertial frame or not, remain pure.

But if we begin to think of quantum measurement outcome
probabilities as not invariant under Lorentz transformations, then
the Von Neumann entropy should not be either. On the other hand
the invariant quantity $\eta_{\mu \nu} \ul{\rho}_{\mu}
\ul{\rho}_{\nu}$ seems a good measure of the mixedness of $\rho$,
an idea which is strongly supported by its equivalent form
(\ref{trace}). With $I(\ul{\rho})$ proportional to the logarithm of $\eta_{\mu
\nu} \ul{\rho}_{\mu} \ul{\rho}_{\nu}$ equation (\ref{information})
becomes:
\begin{align*}
I(\ul{\rho_m})=I(\ul{V_m})+I(\ul{\rho})
\end{align*}
This result is rather interesting as an information conservation
law.

The lines of thought suggested in this last section need to be
anchored in firmer ground and generalized to higher dimensional
quantum systems. Although most
of the mathematical results of this paper stem from the exceptional isomorphism between
$\textrm{Herm}_2^+(\mathbb{C})$ and the future cone of Minkowski
space, there is hope to find a special relativistic
interpretation to $d$ dimensional systems \cite{us}. This is currently
being investigated.  More
generally the authors feel that the correspondence between qubit
quantum operations and special relativity transforms deserves
further attention.

\section{Acknowledgments}
C.E.P would like to thank Gary Gibbons for motivating discussions, EPSRC, the DAMTP, the Cambridge European and
Isaac Newton Trusts for financial support. P.J.A  would like to
thank Anuj Dawar for his patient listening, EPSRC, Marconi, the
Cambridge European  and Isaac Newton Trusts for financial support.

\end{document}